\documentclass{ws-p10x7}

\begin{document}

\title{Predictions for the Decays of Radially-Excited Baryons}

\author{Carl E. Carlson}

\address{Nuclear and Particle Theory Group, Physics Department, College of
William and Mary, Williamsburg, VA 23187-8795, USA\\
E-mail: carlson@physics.wm.edu}

\twocolumn[\maketitle\abstract{
We consider decays of the lowest-lying radially excited  baryons.  Assuming
a single-quark decay approximation,  and negligible configuration mixing, we
make model-independent predictions  for the partial decay widths to final
states with a single meson.  Masses of unobserved states are predicted using
an old mass formula rederived using large-$N_c$ QCD. The momentum dependence of
the one-body decay amplitude is determined phenomenologically by fitting to
observed decays.  Comparison of these predictions to experiment may shed light
on whether the Roper resonance can be interpreted as a  three-quark state. 
}]

\section{Introduction}

I will report results of some studies of excited baryon masses and
decays\cite{largen1,largen2}, concentrating mainly on the radially-excited
baryon multiplet that includes the Roper resonance\cite{largen3}.

Of course, the fundamental QCD degrees of
freedom are quarks and gluons, but we must deal with observed states that are
baryons and mesons.  Our response is to use effective field theory. Here one
first writes down all operators that are consistent with all known symmetries,
and then use some method---in our case large $N_C$---to provide a size estimate
for each operator.  We discard small operators,  keep as many of the large
operators as possible and use them to calculate masses or decay
amplitudes.

To illustrate how effective field theory and large $N_C$ are used, the next
section outlines a modern derivation of the G\"ursey-Radicati\cite{gr} mass
formula.  The result is in itself useful for estimating masses of
undiscovered radially excited baryons.  After that, we show how we make
predictions for decay  widths of radially excited baryons, without assumptions
about spatial wave functions.  

\section{Mass formula}

We look at radially excited baryons where the spatial state, and so also
the spin-flavor state, is totally symmetric.  There are 56 totally symmetric
3-quark states that one can make from $u_\uparrow$, $u_\downarrow$,
$d_\uparrow$,
$d_\downarrow$, $s_\uparrow$, and $s_\downarrow$, where the arrows indicate the
spin projection.  The ground states form the {\bf 56}, and the
radially-excited  states form the {\bf 56$^\prime$}.  The states are the
$N$, $\Lambda$, $\Sigma$, $\Xi$, $\Delta$,
$\Sigma^*$, $\Xi^*$, and $\Omega$.

The mass operators for these states are built from 
the spin $S^i = \sum_\alpha \sigma^i_\alpha / 2$ (the sum is over the
quarks $\alpha$), the flavor operators $T^a = \sum_\alpha \tau^a_\alpha / 2$
(where the $\tau^a$ are a set of 3 $\times$ 3 matrices), and the SU(6)
operators
\begin{equation}
G^{ia} = \sum_\alpha {1\over 2} \sigma^i_\alpha \cdot {1\over 2} \tau^a_\alpha 
\ .
\end{equation}

Terms in mass operators must be rotation symmetric, and
flavor symmetric to leading order.  Not all terms should be included. For
example, in symmetric states matrix elements of $T^2$ and $G^2$ are linearly
dependent on those of
$S^2$ and the unit operator\cite{manrev}.

Flavor symmetry is not exact.  The mass of the strange quark allows non-flavor
symmetric terms in the effective mass operator, visible  as unsummed flavor
indices $a=8$ below.  The effective mass operator is 
\begin{eqnarray}
H_{eff} &=& a_1 1 + {a_2\over N_C} S^2 + \epsilon a_3 T^8 + 
          {\epsilon\over N_C} a_4 S^i G^{i8}  \nonumber \\
         &+& {\epsilon \over N_C^2} a_5 S^2 T^8
          + {\epsilon^2 \over N_C} a_6 T^8 T^8 \nonumber \\
         &+& {\epsilon^2 \over N_C^2} a_7 T^8 S^i G^{i8}
          + {\epsilon^3 \over N_C^2} T^8 T^8 T^8 \ .
\end{eqnarray}

There is an $\epsilon$ for each violation of flavor symmetry, where $\epsilon
\approx 1/3$.  Also, a term that is a product of two or three operators comes
from an interaction that has at least one or two gluon exchanges, and the
strong coupling falls with number of colors as $g^2 \sim 1/N_C$.  (A crucial
theorem is that no perturbation theory diagrams fall slower in $1/N_C$ than the
lowest order ones\cite{manrev}.)

Keeping the first four terms, taking the matrix elements, and
reorganizing leads to
\begin{eqnarray}
M &=& A + B N_s + C [I(I+1) - {1\over 4} N_s^2] \nonumber \\
 &+& D S(S+1)
\end{eqnarray}
where $N_s$ is the number of strange  quarks.  This is the
G\"ursey-Radicati\cite{gr} mass formula.  We use it to predict masses of 4
undiscovered members of the {\bf 56$'$}, given that 4 are known.

\section{The Decays {\bf 56$'$} $\rightarrow$ {\bf56} + meson}

Four of the 8 states in the {\bf 56$'$} are undiscovered or unconfirmed, and
existing measurements have large uncertainty.  However, we anticipate new
results soon from the CLAS detector at CEBAF.  One member of the {\bf
56$'$} is the Roper or N(1440), whose composition has been debated.  Might it
be a qqqG state\cite{hybridroper}, a non-resonant cross section
enhancement\cite{dynamicalroper}, or just a 3-quark radial
excitation\cite{isgurkarl1,glozmanriska,sasaki}?  Our predictions depend upon
the last possibility.

We assume that only single quark operators are needed.  Two
quark operators were studied for decays of orbitally-excited
states\cite{largen2}, and found unnecessary.  There is only one single quark
operator here, so
\begin{equation}
H_{eff} \propto G^{ia} k^i \pi^a \ ,
\end{equation}
where $k^i$ is the meson 3-momentum and $\pi^a$ is a meson field operator.

One gets for the decay widths,
\begin{equation}
\Gamma = {M_f \over 6 \pi M_i} \, k^3 f(k)^2 \, 
     {\sum}  \  | \langle B_f | G_{ja} | B_i \rangle |^2 ,
\end{equation}
where $f(k)$  parameterizes the momentum dependence of the amplitude. For the 7
measured decays it is well fit by $f = (2.8 \pm 0.2)/k$.  With this in hand, we
can predict the widths for 22 decays. The detailed results are
in\cite{largen3}.

To summarize, we have shown how large $N_C$ ideas
provide a modern derivation of the old G\"ursey-Radicati mass formula, and have
predicted decay widths of the {\bf56$'$}.  The success of our predictions would
bolster the view of the Roper as a 3-quark state.

\section*{Acknowledgments}

I thank the conference organizers for their excellent work; Chris
Carone, Jos\'e Goity, and Rich Lebed for pleasant times collaborating; and
the  National Science Foundation for support under Grant No.\ PHY-9900657.

\end{document}